\documentclass[journal]{IEEEtran}

\usepackage{xcolor}
\definecolor{kit-green100}{rgb}{0,.59,.51}
\definecolor{kit-green70}{rgb}{.3,.71,.65}
\definecolor{kit-green50}{rgb}{.50,.79,.75}
\definecolor{kit-green30}{rgb}{.69,.87,.85}
\definecolor{kit-green15}{rgb}{.85,.93,.93}
\definecolor{KITgreen}{rgb}{0,.59,.51}

\definecolor{KITpalegreen}{RGB}{130,190,60}
\colorlet{kit-maigreen100}{KITpalegreen}
\colorlet{kit-maigreen70}{KITpalegreen!70}
\colorlet{kit-maigreen50}{KITpalegreen!50}
\colorlet{kit-maigreen30}{KITpalegreen!30}
\colorlet{kit-maigreen15}{KITpalegreen!15}

\definecolor{KITblue}{rgb}{.27,.39,.66}
\definecolor{kit-blue100}{rgb}{.27,.39,.67}
\definecolor{kit-blue70}{rgb}{.49,.57,.76}
\definecolor{kit-blue50}{rgb}{.64,.69,.83}
\definecolor{kit-blue30}{rgb}{.78,.82,.9}
\definecolor{kit-blue15}{rgb}{.89,.91,.95}

\definecolor{KITyellow}{rgb}{.98,.89,0}
\definecolor{kit-yellow100}{cmyk}{0,.05,1,0}
\definecolor{kit-yellow70}{cmyk}{0,.035,.7,0}
\definecolor{kit-yellow50}{cmyk}{0,.025,.5,0}
\definecolor{kit-yellow30}{cmyk}{0,.015,.3,0}
\definecolor{kit-yellow15}{cmyk}{0,.0075,.15,0}

\definecolor{KITorange}{rgb}{.87,.60,.10}
\definecolor{kit-orange100}{cmyk}{0,.45,1,0}
\definecolor{kit-orange70}{cmyk}{0,.315,.7,0}
\definecolor{kit-orange50}{cmyk}{0,.225,.5,0}
\definecolor{kit-orange30}{cmyk}{0,.135,.3,0}
\definecolor{kit-orange15}{cmyk}{0,.0675,.15,0}

\definecolor{KITred}{rgb}{.63,.13,.13}
\definecolor{kit-red100}{cmyk}{.25,1,1,0}
\definecolor{kit-red70}{cmyk}{.175,.7,.7,0}
\definecolor{kit-red50}{cmyk}{.125,.5,.5,0}
\definecolor{kit-red30}{cmyk}{.075,.3,.3,0}
\definecolor{kit-red15}{cmyk}{.0375,.15,.15,0}

\definecolor{KITpurple}{RGB}{160,0,120}
\colorlet{kit-purple100}{KITpurple}
\colorlet{kit-purple70}{KITpurple!70}
\colorlet{kit-purple50}{KITpurple!50}
\colorlet{kit-purple30}{KITpurple!30}
\colorlet{kit-purple15}{KITpurple!15}

\definecolor{KITcyanblue}{RGB}{80,170,230}
\colorlet{kit-cyanblue100}{KITcyanblue}
\colorlet{kit-cyanblue70}{KITcyanblue!70}
\colorlet{kit-cyanblue50}{KITcyanblue!50}
\colorlet{kit-cyanblue30}{KITcyanblue!30}
\colorlet{kit-cyanblue15}{KITcyanblue!15}

\usepackage{tikz}
\usetikzlibrary{positioning}
\usepackage{pgfplots}
\usepgfplotslibrary{statistics}
\pgfplotsset{compat=1.18}
\usepackage{comment}

\usepackage[nohyperlinks, nolist]{acronym}
\usepackage{booktabs}
\usepackage{siunitx}
\usepackage{bm}
\usepackage{amssymb}
\usepackage{amsmath}
\usepackage[caption=false,font=footnotesize]{subfig}
\usepackage{capt-of}
\usepackage{mathtools}
\usepackage{makecell}

\usepackage[normalem]{ulem} 

\usepackage{titlesec}
\titlespacing*{\subsection}{0pt}{6pt}{3pt}    

\DeclareMathOperator*{\argmax}{\arg\!\max}

\begin{document}
\bstctlcite{IEEEexample:BSTcontrol} 
\newcommand{\e}{\mathrm{e}}
\newcommand{\ihat}{\hat{\imath}}

%
\title{Spiking Neural Network-based Equalization and Demapping in IM/DD Systems: A Comparison}
%
%
%

\author{Eike-Manuel~Edelmann,~\IEEEmembership{Member,~IEEE,}
        Alexander~von~Bank,~\IEEEmembership{Graduate Student Member,~IEEE,}
        and~Laurent~Schmalen,~\IEEEmembership{Fellow,~IEEE}
\thanks{
Parts of this work were presented at the Advanced Photonics Congress: Signal Processing in Photonic Communications (SPPCom) 2023 in paper~\cite{BankSPPCom} and the Neuro-Inspired Computational Elements Conference (NICE) 2025 in paper~\cite{Arnold25nice}. 

This work has received funding from the European Research Council (ERC) under the European Union’s Horizon 2020 research and innovation program (grant agreement No. 101001899).

All authors were with the Communications Engineering Lab (CEL) at Karlsruhe Institute of Technology (KIT), Karlsruhe, Germany at the time this work was carried out. E.-M. Edelmann (e-mail: eike-manuel.edelmann@iis.fraunhofer.de) is now with the Fraunhofer Institute for Integrated Circuits IIS, Erlangen, Germany.
}

}

\markboth{IEEE Journal of Lightwave Technologies,~Vol.~XX, No.~X, Month~20YY}%
{Shell \MakeLowercase{\textit{et al.}}: Bare Demo of IEEEtran.cls for IEEE Journals}
%


\maketitle

\begin{abstract}
In recent years, artificial neural networks (ANNs) have emerged as the de facto standard for addressing challenging problems in communications engineering that are difficult to solve using traditional methods. 
However, the dense matrix multiplications employed in ANNs often result in high computational complexity and, consequently, power-hungry systems.
To overcome these limitations, researchers are turning to alternative computational models, such as spiking neural networks (SNNs), which promise highly energy-efficient computation.
Recent works have shown that SNN-based equalizers and demappers achieve promising results in non-coherent short-reach optical communication systems affected by chromatic dispersion and nonlinear distortion.
However, the proposed approaches differ in several design choices, such as the incorporation of decision feedback, the use of recurrent connections inside the SNN, different neural encodings, and regularization techniques.
A systematic comparison of the various design choices has not yet been reported in the literature.
This article presents a systematic comparison of design choices for SNN-based equalizers and demappers in intensity modulation with direct detection systems, evaluating their impact on bit error rate, spike activity, and model size. 
The results reveal that design and encoding choices affect these metrics differently, leading to inherent trade-offs.
\end{abstract}


\begin{acronym}
    \acro{AWGN}[AWGN]{additive white Gaussian noise}
    \acro{lcd}[LCD]{low chromatic dispersion}
    \acro{hcd}[HCD]{high chromatic dispersion}
    \acro{BER}[BER]{bit error rate}
    \acro{PAM}[PAM]{pulse amplitude modulation}
    \acro{RRC}[RRC]{root-raised-cosine}
    \acro{CD}[CD]{chromatic dispersion}
    \acro{PGU}[PGU]{policy gradient update}
    \acro{SNN}[SNN]{spiking neural network}
    \acro{ANN}[ANN]{artificial neural network}
    \acro{LIF}[LIF]{leaky-integrate-and-fire}
    \acro{ODE}[ODE]{ordinary differential equation}
    \acro{RFE}[RFE]{receptive field encoding}
    \acro{QE}[QE]{quantization encoding}
    \acro{eotm}[EOTM]{end-of-time membrane-potential}
    \acro{NF-SNN}[NF-SNN]{no-feedback \ac{SNN}-based equalizer and demapper}
    \acro{DF-SNN}[DF-SNN]{decision-feedback \ac{SNN}-based equalizer and demapper}
    \acro{BPTT}[BPTT]{backpropagation through time}
    \acro{SG}[SG]{surrogate gradients}
    \acro{wrt}[w.r.t.]{with respect to}
    \acro{RV}[RV]{random variable}
    \acro{RL}[RL]{reinforcement learning}
    \acro{PDF}[PDF]{probability density function}
    \acro{PGT}[PGT]{policy gradient theorem}
    \acro{IMDD}[IM/DD]{intensity modulation with direct detection}
    \acro{CD}[CD]{chromatic dispersion}
    \acro{RRC}[RRC]{root-raised-cosine}
    \acro{MMSE}[MMSE]{minimum mean square error}
    \acro{DFE}[DFE]{decision-feedback equalizer}
    \acro{FIR}[FIR]{finite impulse response}  
    \acro{MaL}[MaL]{machine learning}
    \acro{DL}[DL]{deep learning}
    \acro{ISI}[ISI]{inter-symbol interference}
     \acro{IQR}[IQR]{inter-quartile range}  
    \acro{DSP}[DSP]{digital signal processing}
    \acro{FPGA}[FPGA]{field programmable gate array}
    \acro{IC}[IC]{integrated circuit}
    \acro{STDP}[STDP]{spike-timing-dependent plasticity}
    \acro{DAC}[DAC]{digital-to-analog converter}
    \acro{ADC}[ADC]{analog-to-digital converter}
    \acro{ROP}[ROP]{received optical power}
    \acro{SNR}[SNR]{signal-to-noise ratio}
    \acro{RNN}[RNN]{recurrent neural network}
    \acro{bi-LSTM}[bi-LSTM]{bidirectional long short-term memory}
    \acro{GRU}[GRU]{gated recurrent unit}
    \acro{ASIC}[ASIC]{application-specific integrated circuit}
    
\end{acronym}

\newcommand\bup{\beta_\mathrm{up}}
\newcommand\bdown{\beta_\mathrm{down}}
\newcommand\bcd{\beta_\mathrm{CD}}
\newcommand\disp{\mathrm{ps}\,\mathrm{nm}^{-1}\,\mathrm{km}^{-1}}

\newcommand\ntap{n_\mathrm{tap}}
\newcommand\nff{n_\mathrm{ff}}
\newcommand\nfb{n_\mathrm{fb}}
\newcommand\Nenc{N_\mathrm{enc}}
\newcommand\Nin{N_\mathrm{in}}
\newcommand\Nhid{N_\mathrm{hid}}
\newcommand\Nout{N_\mathrm{out}}
\newcommand\Bt{B_\mathrm{t}}
\newcommand\Be{B_\mathrm{e}}
\newcommand\sn{\sigma_\mathrm{n}^2}

\newcommand\nffir{NF-FIR}
\newcommand\dffir{DF-FIR}
\newcommand\nfann{NF-ANN}
\newcommand\dfann{DF-ANN}
\newcommand\nfsnn{NF-SNN}
\newcommand\dfsnn{DF-SNN}

\newcommand\zavg{S_\mathrm{avg}}

\newcommand\nfe{\ac{NF-SNN}}
\newcommand\dfe{\ac{DF-SNN}}

\newcommand\Zenc{\bm{S}_\mathrm{enc}}
\newcommand\Nt{N_\theta}

\newcommand\bt{\bm{\theta}}
\newcommand\spi{\sigma_{\pi}}

\newcommand\nfrnn{NF-RNN}
\newcommand\dfrnn{DF-RNN}

\definecolor{Rev1green}{rgb}{0, 0, 0}
\definecolor{Rev2red}{rgb}{0, 0, 0}

\section{Introduction}
\IEEEPARstart{T}{elecommunications} are fundamental to today’s connected world, enabling the exchange of information across long distances.
Simultaneously, \ac{MaL} and its subfield \ac{DL} have emerged as groundbreaking technologies, offering innovative solutions to manage the increasing computational complexity of modern communication systems~\cite{Amirabadi25}. 
Typical applications include the mitigation of device and channel impairments using \ac{MaL}~\cite{Carrera21}, where \acp{ANN} serve as the underlying computational model. 
However, the performance of \ac{ANN}-based receivers generally scales with their complexity~\cite[Fig.~1]{Pedro21}, which often results in power-hungry systems on conventional digital electronics due to their reliance on dense matrix multiplications.

In contrast, the human brain employs sparse information processing, activating only a small subset of neurons at any given time, and consumes just tens of watts on average~\cite[p.~664]{Kasabov2018}, achieving energy efficiency that remains unmatched~\cite{Cramer2022}.
Inspired by this, \acp{SNN} aim to replicate the brain’s behavior and efficiency~\cite{Neftci2019surrogate}. 
In \acp{SNN}, neurons and synapses exhibit time-dependent dynamics and communicate via discrete, event-based spikes~\cite{Mass1997}. 
Because spikes are only exchanged when information is processed, the energy consumption of \acp{SNN} can be a fraction of that of comparable \acp{ANN}~\cite{auge2021}. 
Implemented on neuromorphic hardware, \acp{SNN} enable fast, low-power signal processing~\cite[Fig.~1]{ferreiradelima2017neuromorphic}, offering a pathway to energy-efficient \ac{MaL} that combines the computational power of \ac{DL} with the brain’s low-power operation.
\textcolor{Rev1green}{While state-of-the-art digital \ac{MaL} equalizers} 
\textcolor{Rev2red}{ based on recurrent architectures, ranging from simple \acp{RNN} to \ac{bi-LSTM} networks,} 
\textcolor{Rev1green}{achieve high performance, their deployment in optical transceivers is strictly constrained by computational complexity and power consumption. To mitigate these overheads, recent research on digital \ac{MaL} equalizers focuses heavily on algorithmic complexity reduction techniques~\cite{freire_pruning}. These include parameter pruning to remove redundant synaptic connections, weight clustering (or quantization-aware training) to group parameters into shared values, and low-bit quantization to replace floating-point operations with low-precision integer arithmetic~\cite{freire_pruning}. Although these digital optimization strategies substantially reduce the number of active multiply-accumulate operations and memory footprint, the underlying processing hardware remains inherently bound to clock-driven \ac{DSP}. In such synchronous pipelines, dynamic power is continuously consumed across clock cycles to maintain high sampling throughput, regardless of the sparsity or temporal dynamics of the input signal.}

A promising application for SNNs is optical communications, where the mitigation of \ac{CD} and nonlinear distortion remains a key challenge, especially in short-reach fiber-optic systems. 
Recent works~\cite{arnold23journal,BankSPPCom,Arnold25nice} have shown that SNNs can achieve promising results when applied to equalization and demapping in \ac{IMDD} systems affected by \ac{CD} and nonlinear detection.
In~\cite{arnold23journal}, an \ac{SNN}-based equalizer and demapper, which outputs an estimate of the transmit bit sequence based on the most recent received symbols, was proposed.
It outperforms both a linear \ac{MMSE} equalizer and a comparable \ac{ANN}-based equalizer and demapper in terms of \ac{BER}.
In parallel,~\cite{BansSCC} introduced an \ac{SNN}-based equalizer and demapper with decision feedback of the most recently detected symbols.
When applied to links with more severe \ac{ISI}, decision feedback can significantly improve system performance compared to approaches without decision feedback~\cite{BankSPPCom}.

Besides decision feedback, the approaches differ in several aspects. 
First, they employ different neural encodings, which convert real-valued inputs, i.e., received symbols, into binary spike sequences that can be processed by the \ac{SNN}.
Second, the approaches differ in the use of recurrent connections in the hidden layer of the \ac{SNN}, which can enhance performance at the cost of an increased number of model parameters~\cite{BankSPPCom}. 
Third, to reduce the number of spikes required per inference, a regularization term penalizing the average number of spikes per symbol can be applied~\cite{arnold23journal}. 
Finally, the two approaches are evaluated under different channel conditions, which makes a direct comparison difficult.
To date, a systematic comparison of the different design choices is still~missing.

In this paper, we present a systematic comparison of the two approaches, evaluating the impact of the different design choices on system performance.
For the \ac{IMDD} links considered in~\cite{Arnold25nice}, we compare four design aspects: decision feedback versus no decision feedback, recurrent versus non-recurrent architectures, two neural encoding schemes, and the use of regularization.
As a result, 16 different \ac{SNN}-based equalizers and demappers are evaluated for each link. 
For both neural encodings, we further analyze the impact of the encoding parameters on system performance. 
As comparison metrics, we consider the achieved \ac{BER}, the average number of spikes per inference as a proxy of energy-efficiency, and the total number of model parameters.
\section{Spiking Neural Networks}

\subsection{Spiking Neurons}
Spiking neurons form the basic building blocks of \acp{SNN}.
Compared to neurons of traditional \acp{ANN}, spiking neurons fundamentally differ in two aspects: 
first, they posses an internal state and hence exhibit time-dependency.
Second, information is represented and processed in binary events, the so-called \textit{spikes}.

\begin{figure}
    \centering
    \includegraphics[width=.49\textwidth]{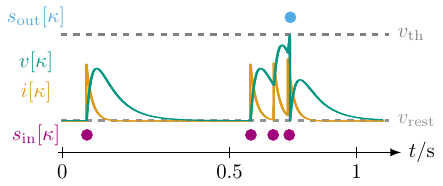}
    \caption{Exemplary temporal dynamics of the \ac{LIF} neuron model using~${\Delta t=\SI{1}{\milli\second}}$, ${\tau_\mathrm{s}=\SI{30}{\milli\second}}$, and ${\tau_\mathrm{m}=\SI{100}{\milli\second}}$.
    Purple dots denote incoming spikes, blue dots denote outgoing spikes.}
    \label{fig:lif_dynamics}
\end{figure}

Similar to the biological neurons they emulate, spiking neurons integrate incoming signals while their state gradually leaks over time, producing spikes that encode information in the timing of their activity. 
More precisely, incoming spikes ${s_\mathrm{in}(t)\in \{0,1\}}$ generate a \textit{synaptic current} ${i(t) \in \mathbb{R}}$, which charges the internal \textit{membrane potential} ${v(t)\in\mathbb{R}}$. 
Simultaneously, leakage continuously discharges $v(t)$. 
When the membrane potential exceeds a predefined \textit{firing threshold} ${v_\mathrm{th} \in \mathbb{R}}$, i.e., when the charging due to the synaptic current overcomes the leakage, an output spike ${s_\mathrm{out}(t)\in \{0,1\}}$ is generated (${s_\mathrm{out}(t)=1}$), and $v(t)$ is reset. 
Fig.~\ref{fig:lif_dynamics} illustrates exemplary temporal dynamics of a spiking neuron.

\subsection{The Leaky-Integrate-and-fire Neuron Model}
Various models of spiking neurons exist, differing in their biological plausability and computational complexity~\cite{izhikevich04}.
A biologically plausible, yet easily computable spiking neuron model is the \ac{LIF} neuron model~\cite{izhikevich04,Neftci2019surrogate}.
The membrane potential $v(t)$ and current $i(t)$ can be described by~\cite{Neftci2019surrogate}
\begin{align} 
    \frac{\mathrm{d}v(t)}{\mathrm{d}t}  &=-\frac{1}{\tau_\text{m}} ((v(t)-v_\mathrm{rest})-i(t))\, , \\[0.5em]
    \frac{\mathrm{d}i(t)}{\mathrm{d}t}  &=-\frac{i(t)}{\tau_\text{s}} + \sum_j w_j s_{\mathrm{in},j}(t) \, ,
\end{align}
where the time constants ${\tau_\text{m}\in\mathbb{R}^+}$ and ${\tau_\text{s}\in\mathbb{R}^+}$ determine the rates at which $v(t)$ and $i(t)$ decay, respectively.
Furthermore, a (trainable) weight ${w_j\in\mathbb{R}}$ modulates the $j$th input spike signal $s_{\mathrm{in},j}(t)$, and $v_\mathrm{rest}$ represents the equilibrium membrane voltage in the absence of input spikes.

Using the forward Euler method, the solution of the \acp{ODE} and therefore the dynamics of the \ac{LIF} neuron model can be approximated~\cite{Cramer2022}.
Assume some arbitrary initial values $v(t_0)$ and $i(t_0)$, with ${t_0=0}$, and ${v_\mathrm{rest}=0}$.
The \acp{ODE} can be solved by numerical integration with a fixed integration step size $\Delta t$, resulting in a system defined at discrete-time instants \mbox{$t=\kappa \Delta t$}, \mbox{$\kappa \in \mathbb{N}$}.
The discrete dynamics of the \ac{LIF} neuron model can be expressed as~\cite{norse}
\begin{align} 
    v[\kappa+1] &= v[\kappa]\cdot \e^{-\frac{\Delta t}{\tau_\text{m}}} + i[\kappa]\cdot \e^{-\frac{\Delta t}{\tau_\text{m}}} \label{eq:lif_v}\\[0.5em]
    i[\kappa+1] &= i[\kappa]\cdot \e^{-\frac{\Delta t}{\tau_\text{s}}} + \sum_j w_j s_{\mathrm{in},j}[\kappa] \label{eq:lif_i} \, ,
\end{align}
where $\Delta t$ is the sampling time of the system and $v[k]$ the membrane potential at time \mbox{$t=\kappa\Delta t$}.
If $v[\kappa]$ exceeds the firing threshold $v_\text{th}$, an output spike $s_\text{out}[\kappa]$ is generated corresponding to 
\begin{align} 
    s_\text{out}[\kappa] = \Theta(v[\kappa]-v_\text{th}) = \begin{cases}
        1\,, \qquad \text{if} \quad v[\kappa]>v_\mathrm{th}\, \\
        0\,, \qquad \text{otherwise} \quad \, ,
    \end{cases} \label{eq:step}
\end{align}
where $\Theta(\cdot)$ denotes the Heaviside step function.
Afterwards, the membrane voltage is reset by \mbox{$v[\kappa] \leftarrow v_\text{rest}$}.  

\subsection{Spiking Neural Networks}
When multiple spiking neurons are interconnected in a directed (possibly cyclic) manner via adaptable weights, an \ac{SNN} is formed. 
Although neurons can be connected arbitrarily, neurons are typically grouped into layers~\cite{Shrestha22survey}.
We distinguish between three different types of layers: the input layer, the hidden layers, and the output layer.
While the input and output layers can be interpreted as the interface of the \ac{SNN}, where external signals are fed to the \ac{SNN} and processed data are read out, the hidden layer processes and transforms the signals.

\begin{figure}
    \centering 
    \includegraphics[width=.49\textwidth]{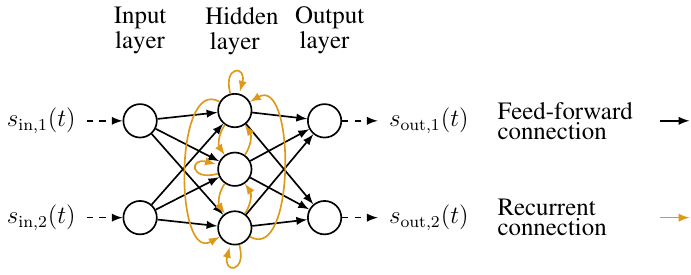}
    \caption{\acs{SNN} with~$\Nin=2$ input neurons,~$\Nhid=3$ hidden neurons, and~$\Nout=2$ output neurons. The hidden layer exhibits recurrent connections.}
    \label{fig:rec_snn}    
\end{figure}

The number of neurons in the different layers is given by~${\Nin\in\mathbb{N}}$ for the input,~${\Nhid\in\mathbb{N}}$ for the hidden, and~${\Nout \in\mathbb{N}}$ for the output layer.
If all layers are connected in a feed-forward manner, a feed-forward \ac{SNN} is obtained~\cite{Schuman17survey}.
In addition, recurrent connections can be incorporated, which can excite or inhibit the firing of the neuron itself and neighboring neurons~\cite{diehl15}.
Fig.~\ref{fig:rec_snn} shows an \ac{SNN} with a single hidden layer and~$\Nin=2$ input neurons,~${\Nhid=3}$ hidden neurons, and~$\Nout=2$ output neurons.
Recurrent connections are highlighted in orange, and feedforward connections in black.

To update the weights of an \ac{SNN}, we use the \ac{BPTT} with \ac{SG} algorithm~\cite{Neftci2019surrogate}. \textcolor{Rev2red}{Since the exact learning mechanisms of the mammalian brain remain incompletely understood, selecting the optimal training rule for \acp{SNN} remains a topic of active debate~\cite{zenke21jointfuture}. Biologically plausible local learning rules, such as \ac{STDP}, rely strictly on the precise relative timing of pre- and post-synaptic spikes. Because they lack a global error signal, purely local rules suffer from local credit assignment limitations and struggle to propagate gradients across multi-layer architectures. As demonstrated in our prior work~\cite[Ch.~2]{edelmann_diss}, restricting STDP training to output layers while using random hidden weights fails when hidden layer activity is low or irregular, ultimately limiting overall equalization performance. In contrast, \ac{BPTT} with \ac{SG} unrolls the computational graph across time, enabling end-to-end optimization of both hidden and output layers within standard machine learning frameworks.} \ac{BPTT} with \ac{SG} is built upon the backpropagation algorithm~\cite{rumelhart1986} and a discrete-time neuron model as given in \eqref{eq:lif_v} and \eqref{eq:lif_i}. At the output layer of the \ac{SNN}, an error term is computed that quantifies the deviation between the true and the desired output. The gradient of the error with respect to the weights is calculated and then backpropagated from the output layer to the input layer, as well as from the most recent simulation time step to all preceding time steps. To overcome the derivative of \eqref{eq:step}, which is zero almost everywhere except at ${v=v_\mathrm{th}}$ and hence hinders backpropagation, the true gradient can be approximated by a surrogate~\cite{Neftci2019surrogate}. For a more detailed description of \ac{BPTT} with \ac{SG}, the interested reader is referred to~\cite{Neftci2019surrogate}.

\subsection{Neural Encoding}\label{sec:neural_decod}
To interface with spiking neurons, real-world signals must first be converted into spike-based representations. 
Before an \ac{SNN} can process a real-valued input~${y\in\mathbb{R}}$, \textit{neural encoding} maps it into spike signals~${s_\mathrm{enc}(t)\in\{0,1\}}$~\cite{auge2021}.Various encoding techniques exist, which encode information either in the rate of spikes~\cite{dayan05}, in the timing of spikes~\cite{auge2021}, or by distributing it across the activity of multiple neurons~\cite{petro20}.
In the following, we will introduce two encoding methods that rely on the distribution of the activity across ${\Nenc\in\mathbb{N}}$ neurons.

\subsubsection{Linear Receptive Field Encoding}
\Ac{RFE} is an encoding method inspired by the visual receptive fields of the human retina~\cite[Ch.~2]{mallot25}. 
It generates $\Nenc$ spike signals, each firing once, where the timing depends on the input~$y$. 
Consequently, different values of $y$ produce unique spike patterns, encoding information in both the firing order and relative timing of the spikes.

To generate the spike pattern, a function ${f_j:\mathbb{R}\rightarrow \mathbb{R}}$ maps each input $y$ to the spike timing ${t_{\mathrm{enc}}^{(j)}\in \mathbb{R}^{+}}$ of the $j$th spike signal, ${j=1,2,\ldots,\Nenc}$.
For the linear \ac{RFE}, $f_j$ is a (piecewise) linear function~\cite{arnold23journal} with 
\begin{align*}
    t_{\mathrm{enc}}^{(j)}(y) = \begin{cases}
        T \cdot \frac{2|y-\mu_j|}{\Delta_j} \, ,  \quad &\text{for}\; \mu_j-\frac{\Delta_j}{2} \leq y \leq \mu_j+\frac{\Delta_j}{2} \, , \\
        T\, , &\text{otherwise}\, ,
    \end{cases}
\end{align*} 
where $T\in\mathbb{R}^{+}$ is the largest possible timing of the spike, and ${\mu_j\in\mathbb{R}}$ and ${\Delta_j \in \mathbb{R}^{+}}$ are two parameters that define the center and the width of the $j$th field.
By applying the floor operator, we obtain the discrete-time encoding 
\begin{align*}
    \kappa_{\mathrm{enc}}^{(j)}(y) =\begin{cases}
        \Big\lfloor (K-1) \frac{2|y-\mu_j|}{\Delta_j} \Big\rfloor \, ,  \; &\mu_j\!-\!\frac{\Delta_j}{2} \leq y \leq \mu_j\!+\!\frac{\Delta_j}{2} \, , \\
        K-1\, , &\text{otherwise}\, ,
    \end{cases}
\end{align*}
where $\kappa_{\mathrm{enc}}^{(j)}\in\{0,1,\ldots,K-1\}$ is the discrete-time spike timing and ${K\in\mathbb{N}}$ the maximum number of simulation steps of the discrete-time \ac{SNN}.
After obtaining $\kappa_{\mathrm{enc}}^{(j)}$, the $j$th spike signal $s_{\mathrm{enc},j}[\kappa]$ is generated by 
\begin{align}
    s_{\mathrm{enc},j}[\kappa] = \begin{cases}
        1\, ,\quad & \text{for}\; \kappa = \kappa_{\mathrm{enc}}^{(j)} \, , \\
        0\, ,&\text{otherwise}\, .
    \end{cases} \label{eq:spike_conv} 
\end{align}

\begin{figure}
    \centering 
    \includegraphics[width=.49\textwidth]{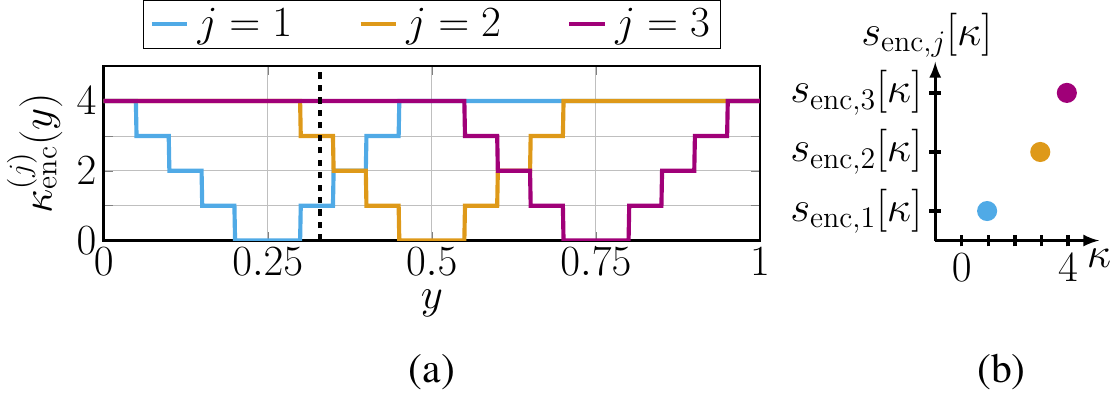}
    \caption{(a) Discrete-time characteristics of a linear \ac{RFE} with~${\Nenc=3}$ fields with~$\mu_j=\frac{j}{4},\,j=1,2,3$, ${\Delta_j=\Delta=\frac12}$ and ${K=10}$. The black dashed line indicates $y=0.33$. \\
    (b) Resulting spike pattern when encoding ${y=0.33}$. }
    \label{fig:lin_field}
\end{figure}
Fig.~\ref{fig:lin_field} illustrates an example of a discrete-time linear \ac{RFE} with ${\Nenc=3}$ fields and the corresponding spike pattern generated when encoding ${y=\num{0.33}}$.

\subsubsection{Quantization Encoding}
\begin{figure}
    \includegraphics[width=.49\textwidth]{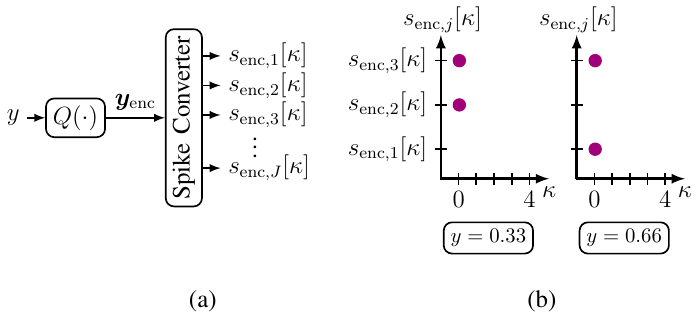}
    \caption{(a) Block diagram of the \ac{QE} with $\Nenc$ output signals. \\
    (b) Resulting spike signals when encoding ${y=0.33}$ and ${y=0.66}$ for~${y_\mathrm{max}=1}$ and ${\Nenc=3}$ are displayed.}
    \label{fig:qe_sketch}
\end{figure}

\Ac{QE} is a encoding technique based on ternary encoding proposed in~\cite{BansSCC}.
It outputs $\Nenc$ spike signals, with the information of $y$ encoded in the subset of spikes that fire at the first time instant, ${\kappa=0}$.
Hence, different values of $y$ produce different subsets of activated spike signals.
Fig.~\ref{fig:qe_sketch}(b) illustrates the resulting spike patterns of two exemplary values of $y$.

The setup of \ac{QE} is displayed in Fig.~\ref{fig:qe_sketch}(a).
Given $y$, it applies a unipolar mid-tread quantizer ${Q:\mathbb{R}\rightarrow \{0,1\}^{\Nenc}}$, which conducts both quantization and bit mapping.
The quantizer characteristic is given by
\begin{align*}
    Q(y) =  
    \begin{cases}
        0\, , & \text{for}\; y<0\, , \\[.5em]
        \Big \lfloor \frac{y}{\Delta_\mathrm{Q}}+\frac12 \Big\rfloor, \, & \text{for}\; ~0<y < (2^{\Nenc}-1)\cdot \Delta_\mathrm{Q}\, ,\\[.5em]
        2^{\Nenc}-1\, , &\mathrm{else}, 
    \end{cases}
\end{align*}
where 
\begin{align*}
    \Delta_\mathrm{Q}=\frac{y_\mathrm{max}}{2^{\Nenc}}
\end{align*}
denotes the step size of the quantizer and~${y_\mathrm{max}\in\mathbb{R}^{+}}$ controls the quantization range.
These quantized integer values are subsequently mapped to a binary sequence~${\bm{y}_\mathrm{enc}\in\{0,1\}^{\Nenc}}$ by using a big-endian binary representation.
Finally, the binary representation is formatted as a set of $\Nenc$ spike signals.
With ${\bm{y}_\mathrm{enc}=(\tilde{y}_{\mathrm{enc},1},\ldots,\tilde{y}_{\mathrm{enc},\Nenc})}$, the spike signals are obtained by 
\begin{align*}
    s_{\mathrm{enc},j}[\kappa]= \begin{cases}
        \tilde{y}_{\mathrm{enc},j}\, , \quad &\text{if}\; \kappa = 0\, ,   \\
        0\, , &\text{otherwise}\, .
    \end{cases}
\end{align*}
Hence, the $j$th bit indicates the spike behavior at the beginning of the $j$th spike signal.

\subsection{Neural Decoding}
To obtain a meaningful output from the \ac{SNN}, \textit{neural decoding} converts the recorded neural activity into a useful representation~\cite{MATHIS24}, such as an estimated class label or a reconstructed signal.
When applied to classification tasks, a promising decoding technique is \ac{eotm} decoding~\cite{BansSCC,Li24}.
Let $\Nout$ denote the number of output neurons, which equals the number of classes in the classification problem, with each neuron corresponding to a single class.
Furthermore, $\hat{\imath}$ denotes the estimate of the class label ${i\in\{1,\ldots,\Nout\}}$.
Then $\hat{\imath}$ is obtained by
\begin{align*}
    \hat{\imath} = \argmax_{i=1,\ldots,\Nout} v_{\mathrm{out},i}[K-1] \, ,
\end{align*} 
which corresponds to reading out and comparing the membrane potential $v_{\mathrm{out},i}[\kappa]$ of the output neurons at the end of the simulation time.

\section{\ac{SNN}-based Equalization and Demapping} 
Novel \ac{SNN}-based algorithms have demonstrated their capability when used for equalization and demapping in non-coherent systems~\cite{arnold23journal,BankSPPCom}.
Recent works proposed two setups for \ac{SNN}-based equalization and demapping: the \textit{\nfe}, which neglects feedback of the most recent transmit symbol estimates~\cite{arnold23journal}, and the \textit{\dfe}, which incorporates the most recent transmit symbol estimates~\cite{BansSCC}.

\begin{figure}
    \centering 
    \includegraphics[width=.42\textwidth]{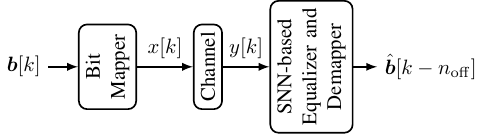}
    \caption{Block diagram of the communication system model.}
    \label{fig:system_model}
\end{figure}

The setup of the discrete-time communication system model we assume is depicted in Fig.~\ref{fig:system_model}.
At time step ${k\in\mathbb{N}}$, the bit sequence ${b[k]\in\{0,1\}^{\log_2(|\mathcal{X}|)}}$ is mapped to a transmit symbol~${x[k]\in\mathcal{X}}$, where ${\mathcal{X}\subset \mathbb{R}}$ denotes the set of transmit symbols\footnote{In general, it is $\mathcal{X}\subset\mathbb{C}$. 
Since this paper focuses on non-coherent systems, we assume $\mathcal{X}\subset\mathbb{R}$.}.
The transmit symbol $x[k]$ is distorted by the channel and ${y[k]\in\mathbb{R}}$ is observed.
The observation is passed to the \ac{SNN}-based equalizer and demapper, which yields an estimate $\hat{\bm{b}}[k-n_\mathrm{off}]$ of the transmitted bit sequence, where ${n_\mathrm{off}\in\mathbb{N}}$ denotes the latency introduced by the equalization process.

\begin{figure*}
    \centering
    \subfloat[\nfe. \label{fig:nfe}]{%
        \begin{minipage}{0.49\textwidth}
            \centering
            \vspace*{-3.8cm}
            \includegraphics[width=\textwidth]{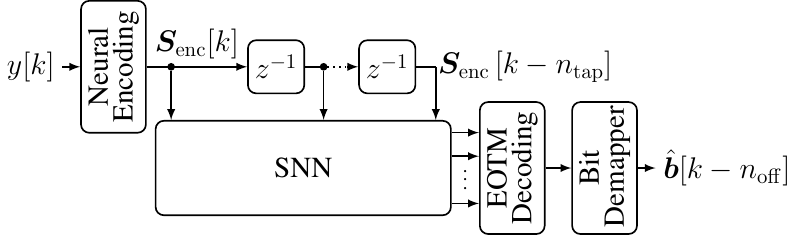}
            \vspace*{.4cm} 
        \end{minipage}
    }%
    \hfill
    \subfloat[\dfe. \label{fig:dfe}]{%
        \includegraphics[width=.49\textwidth]{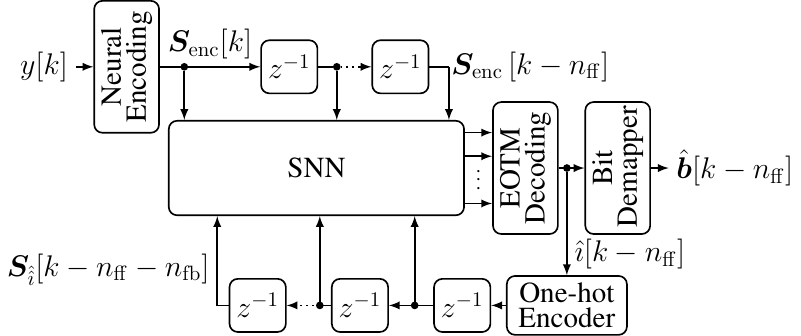}
    }%
    \caption{Block diagram of the \nfe~and \dfe~equalizer and demapper.}%
    \label{fig:equalizers}%
\end{figure*}

\subsection{\nfe~Equalizer and Demapper}
Fig.~\ref{fig:equalizers}(a) depicts the block diagram of the \nfe~equalizer and demapper, denoted only by \nfe~in what follows, with~${\ntap\in\mathbb{N}}$ taps.
By applying the neural encoding, the observation ${y[k]}$ is converted into $\Nenc$ spike signal represented by a binary matrix ${\Zenc[k]\in\{0,1\}^{\Nenc\times K}}$.
The most recent ${\ntap}$ matrices $\Zenc[k]$ are then passed to the \ac{SNN}, thereby providing it with the $\ntap$ most recent received samples $y[k]$. 
After simulating/running the \ac{SNN} for $K$ time steps, \ac{eotm} decoding is applied.
It returns an estimate ${\ihat[k-n_\mathrm{off}]}$ of the class label ${i\in\{1,\ldots,|\mathcal{X}|\}}$, where each class label $i$ corresponds to a transmit symbol ${x\in\mathcal{X}}$. 
A bit demapper converts~${\ihat[k-n_\mathrm{off}]}$ into the estimate ${\hat{\bm{b}}[k-n_\mathrm{off}]}$ of the transmit bit sequence.
Afterwards, the internal states of all neurons in the \ac{SNN} are reset, and the system time of the communication system is increased by ${k \leftarrow k+1}$.

\subsection{\dfe~Equalizer and Demapper}
Fig.~\ref{fig:equalizers}(b) depicts the block diagram of the \dfe~equalizer and demapper, denoted only by \dfsnn~in what follows.
It consists of a feedforward path with ${\nff\in\mathbb{N}}$ taps, which provides the $\nff$ most recent received samples $y[k]$, and a feedback path with ${\nfb\in\mathbb{N}}$ taps, which provides the most recent estimates $\ihat[k]$ to the \ac{SNN}.
As in the \nfe, the feedforward path encodes $y[k]$ into $\Zenc[k]$, delays it by up to ${\nff\in\mathbb{N}}$ steps, and feeds it into the \ac{SNN}.
We then apply \ac{eotm} decoding to obtain the estimate $\ihat[k]$ of the transmit-symbol class label, which is then passed to the bit demapper and to the feedback path. 
In the feedback path, a one-hot encoder converts $\ihat[k]$ into a spike signal, producing a binary matrix~${\bm{S}_{\ihat}[k]\in{0,1}^{|\mathcal{X}|\times K}}$ by
\begin{align*}
    s_{i,\kappa}[k] = \begin{cases}
        1 \, ,\quad&\text{if} \;\ihat[k]=i \; \text{and}\; \kappa=0\, , \\
        0\, , &\text{otherwise}\, , 
    \end{cases}
\end{align*}
where ${s_{i,\kappa}[k]}$ denotes the element in the $i$th row and $\kappa$th column of $\bm{S}_{\ihat}[k]$.
Hence, a single spike indicates the class label given by $\ihat[k]$.

\subsection{Number of Model Parameters}
Based on the architectures of the \nfe~and \dfe, we can compute the number of weights connecting neurons and, consequently, the total number of trainable parameters in the \ac{SNN}.
In this paper, each implementation will use an \ac{SNN} with a single hidden layer to which we can add recurrent connections.
Tab.~\ref{tab:param_calc} displays the number of trainable parameters in dependence on the architecture.
For the \nfe, the number $\Nin$ of input neurons is given by ${\Nin=\ntap\Nenc}$;
For the \dfe, it is given by~${\Nin=\nff\Nenc+\nfb|\mathcal{X}|}$.
Furthermore, for both architectures, the number $\Nout$ of output neurons is given by the number of transmit symbols, i.e.,~$\Nout=|\mathcal{X}|$.

\begin{table}
\vspace{-5mm}
    \centering
    \caption{Number~$\Nt$ of parameters of different architectures.}
    \label{tab:param_calc}
    \begin{tabular}{c p{1.4cm} l}
        \toprule 
        & Recurrent connections   &~$\Nt$ \\
        \midrule
        \nfe & w/o & $\Nhid \cdot (\ntap \Nenc+ |\mathcal{X}|)$ \\
             & w   & $\Nhid \cdot (\ntap \Nenc+\Nhid+|\mathcal{X}|)$ \\[1em]
        \dfe & w/o & $\Nhid\cdot(\nff\Nenc+\nfb|\mathcal{X}|+|\mathcal{X}|)$ \\
             & w   & $\Nhid\cdot(\nff\Nenc+\nfb |\mathcal{X}|+\Nhid+|\mathcal{X}|)$ \\
        \bottomrule
    \end{tabular}
\end{table}

\subsection{Spike Activity Regularization}
The energy efficiency of \acp{SNN} depends on achieving a small number of spike events~\cite{YanBaiWong2024}.
To minimize the number of generated spikes within the \ac{SNN}, a regularization term can be calculated and minimized during optimization~\cite{arnold23journal,Arnold25nice}.
This term depends on both the weights of the \ac{SNN} and the number of spikes generated by each hidden-layer neuron.
Given a batch with~$\Bt$ samples, the regularization loss~${\alpha_\mathrm{r}\in\mathbb{R}^+}$ is calculated by~\cite{Arnold25nice}
\begin{align*}
    \alpha_\mathrm{r} &= \;\alpha_{\mathrm{r},1} \cdot \frac{|| \bt^{(\mathrm{in})}||_2^2}{\Nin\cdot\Nhid}
        + \alpha_{\mathrm{r},2} \cdot \frac{||\bt^{(\mathrm{out})}||_2^2}{\Nhid\cdot\Nout}   \notag \\ 
        &+ \alpha_{\mathrm{r},3} \frac{1}{|\mathcal{H}|}\sum_{j\in\mathcal{H}} \left( \frac{1}{\Bt}  \sum_{m =1}^{\Bt} \left( \alpha_{\mathrm{r},4} - \sum_{\kappa=0}^{K-1} s_{\mathrm{hid},j}^{(m)}[\kappa] \right) \!   \right)^2 \hspace*{-1mm} .
\end{align*} 
The term can be divided into three summands, which are scaled by their respective parameter~$\alpha_{\mathrm{r},n} \in [0,1],\,n=1,2,3$.
The first two summands calculate the mean of the squared weights, where~$\bt^{(\mathrm{in})}$ denotes the vector of weights connecting the input and the hidden layer, and~$\bt^{(\mathrm{out})}$ denotes the vector of weights connecting the hidden and output layer.
The third term returns the deviation of the actual number of generated spikes per hidden neuron from a predefined target~$\alpha_{\mathrm{r},4} \in \mathbb{R}^+$, where~$\mathcal{H}$ is the set of hidden neurons, and~${s_{\mathrm{hid},j}^{(m)}[\kappa] \in\{0,1\}}$ indicates if the~$j$th neuron of the hidden layer emitted a spike at time instant~$\kappa$ when processing the~$m$th sample. 
To apply regularization, the regularization loss $\alpha_\mathrm{r}$ is added to the task loss function during training. 
\def\opac{50}           

\section{Results}
\subsection{Implemented Architectures}
Based on the above considerations, we obtain eight possible architectures for implementing the \ac{SNN}-based equalizer and demapper: \nfsnn~or \dfsnn, with or without recurrent connections, and using either \ac{RFE} or \ac{QE} as the neural encoding scheme.
Furthermore, when training the \ac{SNN}, we can apply or omit regularization, which results in a total of 16 different implementations.
To distinguish between the different approaches, we use the following naming scheme: \texttt{[structure]}$_{\texttt{[enc],[reg],[rec]}}$. 
The parameter ``structure''$\,\in\{$\nfsnn, \dfsnn$\}$ provides the general equalization structure and the parameter ``enc''${\,\in \{\ac{RFE},\,\ac{QE}\}}$ indicates the applied neural encoding.
Furthermore, ``reg''${\,\in\{\varnothing,\mathrm{R}\}}$ indicates if regularization is applied, and ``rec''${\,\in\{\varnothing,\mathrm{rec}\}}$ indicates, if recurrent connections are used.
For instance, \nfsnn$_{\text{RFE},\text{R},\text{rec}}$ denotes the \ac{SNN}-based equalizer without decision feedback, which uses \ac{RFE} as the neural encoding scheme, applies regularization during training, and has recurrent connections.

\subsection{Applied Channels}
\begin{figure*}
    \centering
    \includegraphics[width=\textwidth]{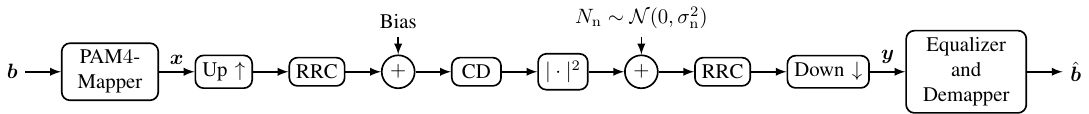}
    \begin{center}
        \caption{System model of the \ac{IMDD} link~\cite{arnold23journal}.}
        \label{fig:imdd_model}
    \end{center}
    \vspace*{-5mm}
\end{figure*}

To compare the implemented architectures, we evaluate them for the transmission of \ac{PAM} symbols over an \ac{IMDD} link.
The continuous-time \ac{IMDD} link can be realistically modeled using discrete-time simulations with oversampling of factor~$\bup$~\cite{arnold_sppcom}.
The resulting model is shown in Fig.~\ref{fig:imdd_model}.
First, the transmit bits are mapped to \ac{PAM} symbols and then upsampled by a factor of $\bup$. 
Pulse shaping is subsequently applied using a \ac{RRC} filter, and a bias is added to ensure that the transmitted signal, normalized to unit power, remains positive when coupled into the fiber.
\ac{CD} within the optical fiber is characterized as an all-pass filter, while the subsequent direct detection at the photodiode is modeled by a square-law operation.
The thermal noise of the receiver is represented as \ac{AWGN} $N_\mathrm{n}$ with noise power $\sn$\textcolor{Rev2red}{$\;\in [-23, -18]~\mathrm{dB}$. To ensure physical validity, this noise range is aligned with the \ac{BER} performance of recent high-speed \ac{IMDD} experimental demonstrations operating at typical received optical power levels between $-8~\mathrm{dBm}$ and $-2~\mathrm{dBm}$~\cite{georg23}.} 
After the matched filtering, the signal is downsampled with $\bdown=\bup$. 
Fiber attenuation is neglected, since the model focuses on dispersion and nonlinear effects, which are the primary performance-limiting factors in short-reach links.

For the simulation of the \ac{IMDD} link, we consider two different sets of channel parameters, which are summarized in Table~\ref{tab:channel_param}.
We refer to the first set of parameters as the \ac{lcd} link, and to the second set of parameters as the \ac{hcd} link.
The parameters are taken from~\cite{Arnold25nice}, and are chosen to represent a realistic short-reach \ac{IMDD} link with low chromatic dispersion and a realistic short-reach \ac{IMDD} link with high chromatic dispersion, respectively. 
\textcolor{Rev1green}{Focusing our evaluation on these two scenarios captures the operational regime strongly dominated by optoelectronic non-linearities, specifically square-law photo-detection at the receiver. 
In~\cite{BansSCC}, we observed that \ac{SNN}-based, \ac{ANN}-based, and \ac{FIR} filter-based \acp{DFE} exhibit similar performance when operating over purely linear channel models (e.g., the Proakis baselines~\cite[p.~654]{proakis2008}).
In contrast, non-linear \ac{IMDD} links represent the critical regime where \acp{SNN} and \acp{ANN} leverage their non-linear mapping capabilities~\cite[Fig.~2]{BankSPPCom}. Furthermore, prior investigations across varying link lengths~\cite[Fig.~3]{BankSPPCom} confirm that \acp{SNN} reliably track the performance trends of conventional \acp{ANN}. Provided an appropriate neural encoding scheme is selected, we believe that \acp{SNN} offer a general-purpose framework capable of replacing standard \ac{ANN}-based \acp{DSP} blocks across diverse optical communication links.}

\begin{table}
    \centering
    \caption{Parameter set of the simulated \ac{IMDD} links.}
    \label{tab:channel_param}
    \begin{tabular}{lcc}
        \toprule 
         & \acs{lcd} link & \acs{hcd} link \\
        \midrule 
        Set of transmit symbols & $\{-3,-1,1,3\}$ & $\{0,1,\sqrt{2},\sqrt{3}\}$ \\
        Upsampling factor $\bup$ & 3 & 3 \\
        Baudrate $R_\mathrm{sym}$ & $112\,\mathrm{GBd}$ & $50\,\mathrm{GBd}$ \\
        Wavelength $\lambda$ & $\SI{1270}{\nano\meter}$ & $\SI{1550}{\nano\meter}$ \\
        \makecell[l]{Dispersion \\ coefficient $\bcd$} & $-5\,\disp$ &  $-17\,\disp$ \\
        Fiber length $L_\mathrm{CD}$ & $\SI{4}{\kilo\meter}$ & $\SI{5}{\kilo\meter}$ \\ 
        Roll-off factor $\beta_\mathrm{RRC}$ & $\num{0.2}$ & $\num{0.2}$ \\
        Bias & 2.25 & 0.25 \\
        \bottomrule
    \end{tabular}
\end{table}

\subsection{Comparison Metrics}
To compare the various architectures, we use three different metrics:
The \ac{BER} to evaluate the performance of the obtained equalizer and demapper when applied in the given links.
The second metric is the number ${\zavg\in\mathbb{R}^{+}}$ of generated spikes in the hidden layer, which serves as a proxy for the energy-consumption per symbol decision\textcolor{Rev1green}{, which captures the foundational algorithmic advantage of \acp{SNN}, where dynamic power consumption is fundamentally activity-gated~\cite{dampfhoffer2022}}.
It is obtained by 
\begin{align*}
    \zavg = \frac{1}{\Be} \sum_{m =1}^{\Be} \sum_{j\in\mathcal{H}} \sum_{\kappa=0}^{K-1} s_{\mathrm{hid},j}^{(m)}[\kappa] \, ,
\end{align*}
where $s_{\mathrm{hid},j}^{(m)}[\kappa]\in\{0,1\}$ denotes the spike emitted by the $j$th neuron of the hidden layer at time instant $\kappa$ when processing the $m$th sample, and $\Be$ is the number of samples used for evaluation.
As a third metric, we use the number $\Nt$ of trainable parameters, which is directly linked to the required memory in digital systems or, alternatively, the physical footprint and hardware resource requirements in analog implementations.

After evaluating the different architectures with respect to these three metrics, we compare the best-performing \ac{SNN}-based equalizer and demapper with benchmark receivers for the \ac{lcd} and \ac{hcd} links, respectively.
As benchmarks, we employ the linear \ac{MMSE} equalizer~\cite[Sec.~9.4]{proakis2008} and the \ac{DFE}~\cite[Sec.~9.5]{proakis2008}, both implemented using \ac{FIR} filters.
A Lloyd-Max quantizer~\cite{LLoyd_max} at the output of both equalizers serves as optimized hard-decision detector of the received symbols.
We refer to these approaches in short as \nffir~ and \dffir. 
Additionally, we implement \ac{ANN}-based equalizers and demappers that follow the same structure as the \ac{SNN}-based counterparts, but with neurons replaced by ReLU-activated units, and without neural encoding.
These approaches are denoted in short as \nfann~ and \dfann.
\textcolor{Rev1green}{To isolate the impact of the neural encoding, we also feed the \dfann~with \ac{QE}. Since conventional \acp{ANN} lack a temporal dimension, we input only the first time step of the \ac{QE} representation, which already carries the complete quantized input information.} 
In the comparison, the benchmark equalizers employ the same number of taps as the SNN-based equalizers.

\subsection{Hyperparameters}
\begin{table}
    \centering
    \caption{Link- and architecture-specific parameters of the \ac{SNN}-based equalizers and demappers.}
    \label{tab:eq_param}
    \begin{tabular}{lcc}
        \toprule
        & \acs{lcd} link & \acs{hcd} link \\
        \midrule 
        Number of equalizer taps $\ntap$ & 7 & 21\\
        Number of feedforward taps $\nff$ & 4 & 11 \\
        Number of feedback taps $\nfb$ & 3 & 10 \\
        Number of hidden neurons $\Nhid$ & 40 & 60 \\
        \bottomrule
    \end{tabular}
\end{table}

\begin{figure*}[t!]
    \centering 
    \subfloat[Fixed $\Nenc=8$. \label{fig:lcd_compare_K}]{%
        \centering
        \begin{minipage}[]{.44\textwidth}
            \includegraphics[width=\textwidth]{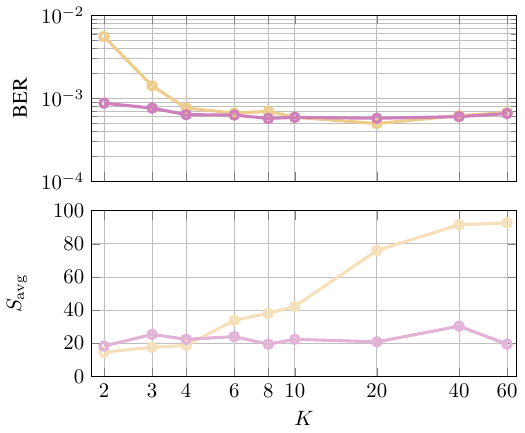}
        \end{minipage}
    }%
    \hspace*{1cm} 
    \subfloat[Fixed $K=6$.\label{fig:lcd_compare_n_enc}]{%
        \centering
        \begin{minipage}[]{.44\textwidth}
            \includegraphics[width=.975\textwidth]{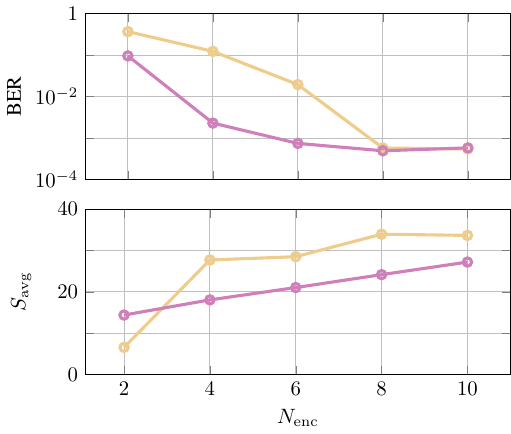}
            \vspace{-.8mm}
        \end{minipage}
    }

    \begin{minipage}[b]{\textwidth}
        \centering
        \includegraphics[width=.3\textwidth]{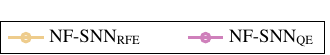}
    \end{minipage}
    
    \caption{\Ac{BER} and $\zavg$ as functions of the encoding parameters $K$ and $\Nenc$, evaluated for the \nfsnn~with \ac{RFE} and \ac{QE} encoding schemes, without regularization or recurrent connections, for the \ac{lcd} link at $\sn=\SI{-20}{\decibel}$.}
    \label{fig:lcd_compare}
    \vspace*{-5mm}
\end{figure*}

The link- and architecture-specific parameters of all \ac{SNN}-based equalizers and demappers are chosen as shown in Tab.~\ref{tab:eq_param} and are based on~\cite{Arnold25nice}. \textcolor{Rev1green}{A more detailed impact analysis of these hyperparameters is provided in~\cite[Ch.~4]{edelmann_diss}, where we demonstrated that the chosen parameters offer a good trade-off between performance and complexity. For instance, decreasing the batch size degrades performance, whereas further increasing it does not yield additional improvement~\cite[Fig.~4.9(a)]{edelmann_diss}.}
Furthermore, training is carried out at a fixed noise power of $\sn=\SI{-20}{\decibel}$, which has been shown to be favorable compared to training with varying noise power~\textcolor{Rev1green}{\cite[Fig.~2]{BankSPPCom}~\cite[Fig.~4.9(b)]{edelmann_diss}}.
The models are trained using the cross-entropy loss, where the softmax-function is applied to the membrane potentials of the output layer to obtain the predicted probabilities of the transmit symbols.
During evaluation, \ac{eotm} as described in Sec.~\ref{sec:neural_decod} is used.
A batch size of ${\Bt=5\cdot 10^4}$ is used during training, while ${\Be=10^7}$ samples are used for evaluation.
Regularization parameters are set to $\alpha_{\mathrm{r},1}=\alpha_{\mathrm{r},2}=10^{-4}$, $\alpha_{\mathrm{r},3}=5\cdot 10^{-4}$, and $\alpha_{\mathrm{r},4}=\num{0.5}$~\cite{Arnold25nice}.
Optimization is performed using the Adam optimizer with a learning rate of $10^{-3}$ for~${10^5}$ epochs.

Various values of $K$ and $\Nenc$ are considered, and the best-performing values are chosen for the final comparison. 
When using \ac{RFE} as the encoding scheme, the centers of the receptive fields are initialized as ${\mu_j = \frac{j}{\Nenc} , y_{\mathrm{max}}}$, for ${j=1,\ldots,\Nenc}$, with ${y_\mathrm{max}=7}$. 
The widths of the fields are chosen uniformly as ${\Delta_j = \Delta = 2}$.
If \ac{QE} is used, we also choose $y_\text{max} = \num{7}$.

All \ac{LIF} neurons in the hidden and output layers are initialized with ${\tau_\text{s}=\SI{5}{\milli\second}}$.
Additionally, ${\tau_\text{m}=\SI{10}{\milli\second}}$ is used in the hidden layer, whereas ${\tau_\text{m}=\SI{1}{\second}}$ is used in the output layer.
All \acp{SNN} are implemented and optimized using \texttt{Norse}~\cite{norse}, which uses the \ac{BPTT} with \ac{SG} algorithm for optimization.
\textcolor{Rev2red}{
Note that while time constants are specified in milliseconds following standard software simulation conventions, the underlying system equations are dimensionless with respect to the symbol duration. When deploying our proposed \ac{SNN}-based equalizers on custom communication-centric neuromorphic \acp{ASIC}, such as the SENNA architecture, nanosecond dynamics enable an ultra-low response time of $20\,\mathrm{ns}$~\cite{pscheidl2025senna}.}

\subsection{Sensitivity and Robustness Analysis}
\subsubsection{Influence of Neural Encoding Parameters}

We first analyze the effect of the encoding parameters $K$ and $\Nenc$ by optimizing and evaluating \nfsnn$\text{QE}$ and \nfsnn$\text{RFE}$ for the \ac{lcd} link at $\sn=\SI{-20}{\decibel}$. 
In Fig.~\ref{fig:lcd_compare}(a), we vary $K$ with $\Nenc=8$, and in Fig.~\ref{fig:lcd_compare}(b), we vary $\Nenc$ with~$K=6$. \newpage

Using \ac{RFE} as the encoding scheme, the \ac{BER} remains unchanged down to~${K=6}$, see Fig.~\ref{fig:lcd_compare}(a).
Further decreasing $K$ results in a significant increase in the \ac{BER}.
With decreasing $K$, $\zavg$ also decreases, with ${\zavg\approx90}$ for ${K=60}$ down to ${\zavg\approx 30}$ for ${K=6}$.
When using \ac{QE} as the encoding scheme, the figures suggest that both the \ac{BER} and $\zavg$ are independent of $K$.   
Since \ac{QE} encodes all information in the spike pattern at the first time instant, i.e.,~$\kappa=0$, the \ac{SNN} can directly access and process all relevant information.
Hence, adding more time instants does not affect the \ac{BER} or $\zavg$.
In contrast, \ac{RFE} encodes the information in the order and relative timing of the spikes.
Hence, choosing $K$ too low results in limited resolution.
Consequently, with decreasing $K$, the \ac{BER} drastically increases.
Since $K$ is directly linked to the number of time steps the \ac{SNN} is simulated, we opt to choose $K$ as low as possible.
From Fig.~\ref{fig:lcd_compare}(a), we can conclude that for \ac{RFE}, $K=6$ is a good choice, as it achieves a low \ac{BER} while keeping $\zavg$ low.
To ensure a fair comparison of both encoding schemes, we fix $K=6$ for \ac{QE} and \ac{RFE} in all subsequent simulations.

When decreasing $\Nenc$, see Fig.~\ref{fig:lcd_compare}(b), for both encoding schemes and ${\Nenc<8}$, the \ac{BER} deteriorates, and $\zavg$ decreases.
For \ac{RFE}, the deterioration is more significant than for \ac{QE}.
To ensure a low \ac{BER}, for further comparisons, we fix~$\Nenc=8$.

\subsubsection{Impact of Regularization}
\begin{figure}
        \centering
        \includegraphics[width=.45\textwidth]{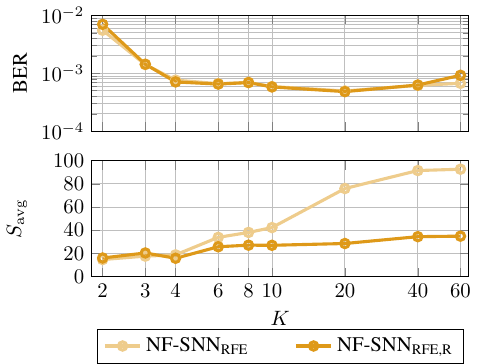}
        \caption{Impact of regularization for the \nfsnn$_\mathrm{RFE}$ with fixed $\Nenc=8$ on the \ac{lcd} link.}
        \label{fig:lcd_compare_K_reg}
\end{figure}
To illustrate the effect of regularization, Fig.~\ref{fig:lcd_compare_K_reg} compares \nfsnn$_{\text{RFE}}$ and \nfsnn$_{\text{RFE,R}}$ for $\Nenc=8$ across different values of $K$ on the \ac{lcd} link..
While the effect of regularization on the \ac{BER} seems negligible, it significantly reduces $\zavg$ for~${K>4}$.
We conclude that regularization is a powerful tool to reduce hidden layer spike activity, and hence the energy consumption of \ac{SNN}-based equalizers and demappers, while maintaining a low \ac{BER}.

\subsubsection{Statistical Significance Across Runs}
\begin{figure}
    \centering 
    \begin{minipage}{0.49\textwidth}
        \centering
        \includegraphics[width=\textwidth]{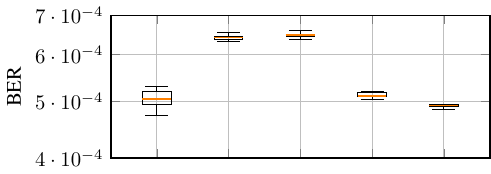}
    \end{minipage}%

    \begin{minipage}{0.49\textwidth}
        \centering
        \hspace*{2.24mm}
        \includegraphics[width=.958\textwidth]{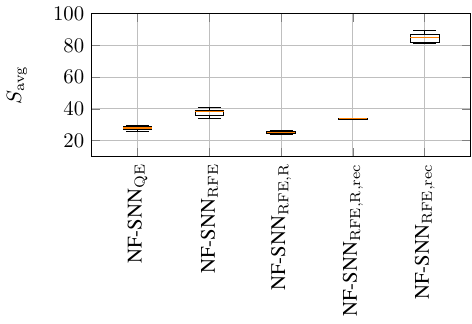}
    \end{minipage}
    
    \caption{Boxplot of \ac{BER} and~$\zavg$ of six different \ac{SNN}-based equalizers and demappers for the \ac{lcd} link. To obtain each box, independent initializations of the respective approach were optimized ten times, and evaluated for~${\sn=-\SI{20}{\decibel}}$.}
    \label{fig:lcd_boxplot}
\end{figure}

In Fig.~\ref{fig:lcd_boxplot}, we investigate the statistical significance of the results across different runs.
The boxplots show the distribution of the \ac{BER} and $\zavg$ for five different setups of the \ac{SNN}-based equalizer and demapper.
For each of the displayed architectures, we independently initialized the model parameters ten times, optimized each initialization separately, and then evaluated at~$\sn=\SI{-20}{\decibel}$.
The orange line denotes the median, the box spans the lower and upper quantiles, and the whiskers extend to at most 1.5 times the interquartile range.
From Fig.~\ref{fig:lcd_boxplot}, we can conclude that a single initialization and optimization is statistically significant across different runs, as the intervals of the achieved \ac{BER} and $\zavg$ for the different architectures are very small.

\subsection{Results: \ac{lcd} link}
\begin{figure}

    \begin{minipage}{\columnwidth}
        \centering
        \includegraphics[width=\textwidth]{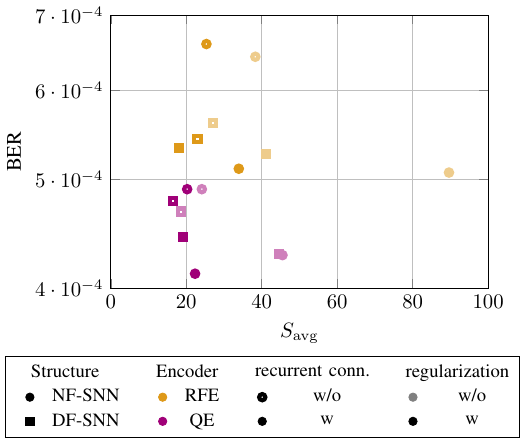}
        \captionof{figure}{Comparions of several SNN-based equalizers and demappers for the \ac{lcd} link.}
        \label{fig:lcd_scatterplot}
    \end{minipage}

    \vspace*{.5cm} 

    \begin{minipage}{\columnwidth}
        \centering
        \captionof{table}{Number $\Nt$ of parameters for the \ac{lcd} link and different architectures.}
        \label{tab:param_lcd}
        \begin{tabular}{ccc}
            \toprule 
                & \nfsnn & \dfsnn \\
            \midrule
            w/o rec. conn. & $2\,400$ & $1\,920$ \\
            w rec. conn.  & $4\,000$ & $3\,520$ \\
            \bottomrule
        \end{tabular}
    \end{minipage}

    \vspace*{.5cm}

    \begin{minipage}{\columnwidth}
        \centering
        \includegraphics[width=\textwidth]{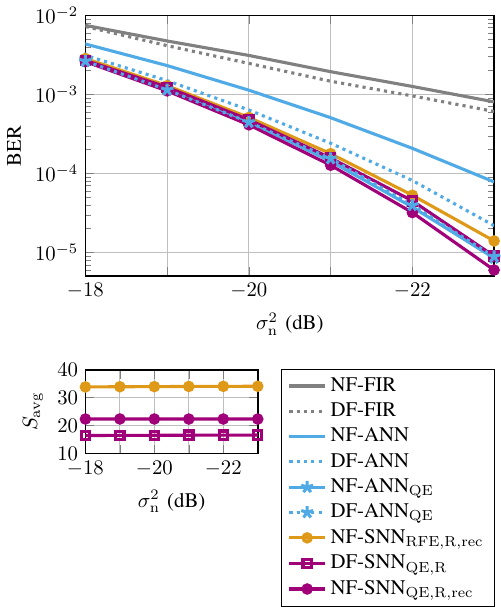}
        \captionof{figure}{Comparison of the best \ac{SNN}-based equalizers and demappers with benchmark receivers for the \ac{lcd} link.}
        \label{fig:lcd_ber}
    \end{minipage}
\end{figure}
Fig.~\ref{fig:lcd_scatterplot} compares the \ac{BER} and $\zavg$ of all 16 implementations of the \ac{SNN}-based equalizer and demapper for the \ac{lcd} link when evaluated for ${\sn=\SI{-20}{\decibel}}$.
Square markers indicate the \nfsnn, whereas circular markers indicate the \dfsnn. 
Colors represent the applied encoding scheme. 
Marker transparency indicates whether regularization is applied, while the presence of recurrent connections is indicated by solid or hollow markers.
Note that the range of the y-axis is relatively small, and thus all implementations achieve a similar \ac{BER} performance.

We observe that \ac{QE} generally yields a smaller \ac{BER} than approaches using \ac{RFE}.
In terms of $\zavg$, no systematic differences can be observed.
With regularization, $\zavg$ is reduced across all approaches, with a negligible effect on the \ac{BER}.
Adding recurrent connections results in minor improvements in terms of \ac{BER}.
However, without regularization, $\zavg$ increases.
From the figure, we identify the \nfsnn$_{\text{QE,R,rec}}$ as the best performing approach, as it achieves the lowest \ac{BER}.

In Tab.~\ref{tab:param_lcd}, we display the number $\Nt$ of model parameters for the \ac{lcd} link and $\Nenc=8$.
Owing to the one-hot encoding in the feedback path of the \dfsnn~(see Fig.~\ref{fig:equalizers}(b)), the \dfsnn~contains ${\nfb(\Nenc-|\mathcal{X}|)\cdot\Nhid=480}$ fewer parameters than the \nfsnn.
Adding recurrent connections, $\Nhid^2$ additional parameters are added, which results in an increase of $\Nt$ by $1\,600$.
Consequently, to reduce $\Nt$, a \dfsnn~architecture without recurrent connections is preferred for low-complexity implementations.
Based on~Tab.~\ref{tab:param_lcd} and Fig.~\ref{fig:lcd_scatterplot}, we identify the \dfsnn$_{\text{QE,R}}$ as the approach with the lowest $\Nt$ and $\zavg$. 

In Fig.~\ref{fig:lcd_ber}, we compare the best-performing \ac{SNN}-based equalizers and demappers with the benchmark receivers for different values of $\sn$.
We plot the achieved \ac{BER} and $\zavg$ of \nfsnn$_{\text{QE,R,rec}}$, which achieves the best \ac{BER} performance, and \dfsnn$_{\text{QE,R}}$, which performs best in terms of $\zavg$ and~$\Nt$.
We also include the \nfsnn$_{\mathrm{RFE,R,rec}}$, as it is the \ac{RFE}-based approach that achieves the lowest \ac{BER}.

All three \ac{SNN}-based equalizers and demappers outperform the benchmark receivers \textcolor{Rev1green}{without neural encoding} across the considered range of $\sn$.
When compared to the \dfann, which is the best benchmark, the \ac{SNN}-based approaches achieve a significantly smaller \ac{BER} at low $\sn$.
At a \ac{BER} of $10^{-3}$, the \nfsnn$_{\text{QE,R,rec}}$ achieves a gain of approximately $\SI{0.7}{\decibel}$ when compared to the \dfann.

We attribute the superiority of the \ac{SNN}-based approaches to the high-dimensional input embedding provided by the neural encoding. 
Since the \ac{ANN} connects each scalar input to $\Nhid$ hidden neurons, the interaction is restricted to a simple linear weighting at the input stage. 
In contrast, the \ac{SNN} encoding expands each measurement into an $\Nenc$-dimensional space, effectively providing $\Nenc\cdot \Nhid$ degrees of freedom per input tap.  
This interpretation is supported by \textcolor{Rev1green}{the \ac{BER} curves of \nfann$_{\mathrm{QE}}$ and \dfann$_{\mathrm{QE}}$, which combine \ac{QE} with \nfann~and \dfann, respectively. Both achieve performance comparable to the \ac{SNN}-based equalizers, thereby outperforming their unencoded counterparts.} 
Consequently, we recognize that the performance gain of the \acp{SNN} is \textcolor{Rev1green}{to some extent} driven by the high-dimensional neural encoding. 
However, it is important to emphasize that such encoding is an inherent and necessary component of \acp{SNN} to enable sparse, event-driven processing, whereas for \acp{ANN}, this expansion would massively increase computational complexity and, consequently, power consumption.

While the \nfsnn$_{\text{QE,R,rec}}$ is the best performing approach in terms of \ac{BER}, the \dfsnn$_{\text{QE,R}}$ minimizes both $\zavg$ and $\Nt$ at the cost of a minor \ac{BER} degradation.
From the figures, we can further conclude that \ac{SNN}-based equalizers and demappers optimized at a fixed $\sn$ generalize well across a wide range of $\sn$, which is aligned with our findings from~\cite{BankSPPCom}.
Moreover, $\zavg$ remains approximately constant over the considered range of~$\sn$.

From Fig.~\ref{fig:lcd_ber}, we conclude that for the \ac{lcd} link, the proposed \ac{SNN}-based equalizers and demappers outperform the benchmark receivers in terms of \ac{BER}.
While all \ac{SNN}-based approaches achieve a similar \ac{BER}, they greatly differ in terms of $\zavg$ and $\Nt$.

\subsection{Results: \ac{hcd} Link}
\begin{figure}
    \begin{minipage}{\columnwidth}
            \centering
            \includegraphics[width=\textwidth]{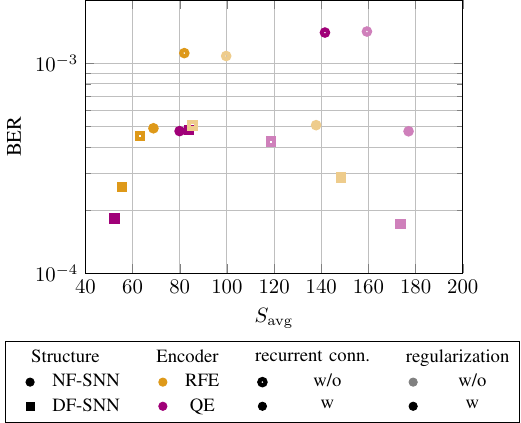}
            \captionof{figure}{Comparison of several SNN-based equalizers and demappers for the \ac{hcd} link.}
            \label{fig:hcd_scatterplot}
    \end{minipage}

    \vspace*{.5cm}

    \begin{minipage}{\columnwidth}
        \centering
        \captionof{table}{Number $\Nt$ of parameters for the \ac{hcd} link and different architectures.}
        \label{tab:param_hcd}
        \begin{tabular}{ccc}
            \toprule 
                & \nfsnn & \dfsnn \\
            \midrule
            w/o rec. conn. & $10\,320$ & $7\,920$ \\
            w rec. conn.  & $13\,920$ & $11\,520$ \\
            \bottomrule
        \end{tabular}
    \end{minipage}

    \vspace*{.5cm}

    \begin{minipage}{\columnwidth}
        \centering
        \includegraphics[width=\textwidth]{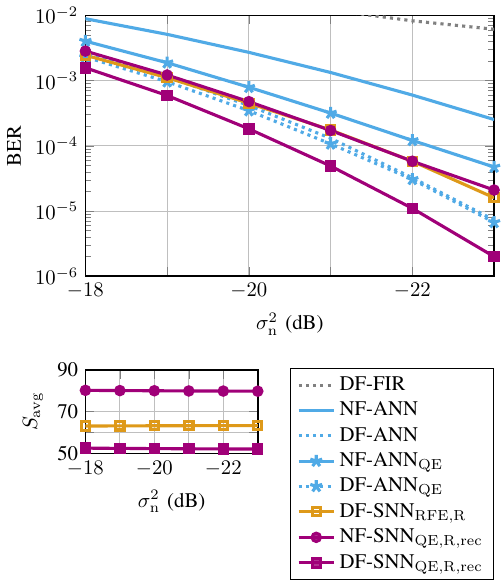}
        \captionof{figure}{Comparison of the best \ac{SNN}-based equalizers and demappers with benchmark receivers for the \ac{hcd} link.}
        \label{fig:hcd_ber}
    \end{minipage}

\end{figure}
Next, we conduct the same analysis for the \ac{hcd} link.
Again, 16 different implementations of the \ac{SNN}-based equalizer and demapper are evaluated.
Their achieved \ac{BER} and $\zavg$ at~$\sn=\SI{-20}{\decibel}$ are shown in Fig.~\ref{fig:hcd_scatterplot}.   
Note that the range of the y-axis is larger than for the \ac{lcd} link, which indicates that for the \ac{hcd} link, the performance heavily depends on the implementation.
Furthermore, for all implementations, we observe a larger $\zavg$ when compared to the \ac{lcd} link.
In general, \dfsnn-based implementations outperform \nfsnn-based implementations in terms of \ac{BER}.
While all \nfsnn-based implementations are not capable of achieving a \ac{BER} smaller than ${4\cdot 10^{-4}}$, some \dfsnn-based implementations achieve a \ac{BER} below ${2\cdot 10^{-4}}$.
We conclude that for the \ac{hcd} link, the incorporation of decision feedback improves performance.
When adding recurrent connections, all approaches achieve a smaller \ac{BER} than their counterparts without recurrent connections.
When applying regularization, $\zavg$ is reduced for all approaches.
Lastly, when comparing the neural encoding scheme, we cannot identify a clear trend in terms of \ac{BER}. 
From the figure, we identify the \dfsnn$_{\text{QE,R,rec}}$ as the best-performing approach in terms of \ac{BER} and $\zavg$.

Tab.~\ref{tab:param_hcd} shows the number $\Nt$ of model parameters for the \ac{hcd} link and $\Nenc=8$.
As for the \ac{lcd} link, architectures with decision feedback exhibit fewer parameters than \nfsnn-based structures, and adding recurrent connections increases $\Nt$.
Among all \dfsnn-based approaches without recurrent connections, i.e., those with the smallest $\Nt$, we identify \dfsnn$_{\text{RFE,R}}$ as the best-performing approach, see Fig.~\ref{fig:hcd_scatterplot}.
We furthermore identify \nfsnn$_\text{QE,R,rec}$ as the best performing \nfsnn-based approach.

In Fig.~\ref{fig:hcd_ber}, we compare these three approaches with the benchmark receivers for different values of $\sn$.
Note that the \nffir~only achieves \ac{BER} values above $10^{-2}$, and is hence not shown.
In general, for the \ac{hcd} link, incorporating decision feedback significantly improves performance; compare \dfann~with \nfann, and see Fig.~\ref{fig:hcd_scatterplot}.
The \dfsnn$_\text{QE,R,rec}$ achieves the best \ac{BER} performance, clearly outperforming all other equalizers.
At a \ac{BER} of $10^{-3}$, the \dfsnn$_\text{QE,R,rec}$ achieves a gain of approximately $\SI{1}{\decibel}$ when compared to the \dfann.
If decision feedback is omitted, we notice a clear degradation in performance, compare the \nfsnn$_\text{QE,R,rec}$ and \dfsnn$_\text{QE,R,rec}$, emphasizing the importance of decision feedback for the given link.
When using \dfsnn$_{\text{RFE,R}}$, which requires only $68.75\%$ of the parameters of \dfsnn$_{\text{QE,R,rec}}$ since no recurrent connections are used, the performance also noticeably degrades, achieving a performance similar to the \nfsnn$_\text{QE,R,rec}$.
We want to emphasize that the \ac{SNN}-based implementations clearly outperform their \ac{ANN}-based counterparts, compare \nfsnn$_\text{RFE,R,rec}$ and \nfann, as well as \dfsnn$_\text{QE,R,rec}$ and \dfann.
\textcolor{Rev1green}{Finally, comparing \dfann~with \dfann$_\mathrm{QE}$ shows similar performance, whereas introducing the neural encoding for \nfann~yields a clear performance improvement. This indicates that while neural encoding can enhance \ac{ANN}-based architectures under certain conditions, its benefits do not generalize to all feedforward models. We furthermore conclude that for the \ac{hcd} link, the superior performance of \ac{SNN}-based equalizers stems intrinsically from the temporal processing intrinsic to \acp{SNN} rather than solely from the input encoding.} 

\textcolor{Rev2red}{To investigate the interaction of neural encoding and temporal processing in more detail, we further supply a vanilla \ac{RNN} baseline and evaluated its performance on the \ac{hcd} link. For a fair comparison against \ac{SNN}-based equalizers, we also applied \ac{QE} prior to the \ac{RNN}-based equalizers. While \nfrnn$_\mathrm{QE}$~denotes the \ac{RNN}-based equalizer without decision feedback and with \ac{QE}, \dfrnn$_\mathrm{QE}$~ denotes the \ac{RNN}-based equalizer with decision feedback and with \ac{QE}.}

\begin{figure}
    \centering
    \includegraphics[width=.49\textwidth]{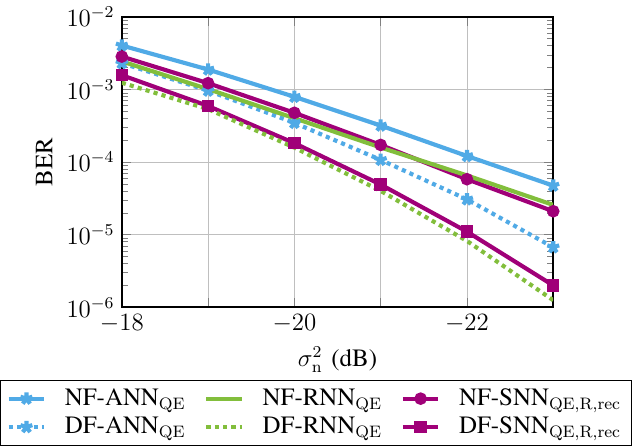}
    \caption{\textcolor{Rev2red}{Comparison of \ac{ANN}-, \ac{RNN}-, and \ac{SNN}-based equalizers with \ac{QE} for the \ac{hcd} link.}}
    \label{fig:hcd_rnn}
\end{figure}

\textcolor{Rev2red}{The results are displayed in Fig.~\ref{fig:hcd_rnn}.
Both \ac{RNN}-based equalizers outperform their \ac{ANN}-based counterparts and perform similar to their \ac{SNN}-based counterparts. This finding provides a crucial technical insight: feedforward \acp{ANN} struggle with the neural encoding not due to a general limitation of artificial neural networks, but because static architectures must map the high-dimensional spike burst in a single pass, whereas recurrent architectures can process and resolve the encoded feature state across multiple internal state transitions. In contrast, architectures with temporal depth (such as \acp{RNN} and \acp{SNN}) are inherently capable of decoding these temporal dynamics to effectively mitigate severe channel \ac{ISI}.}

\subsection{Discussion}
For both investigated links, the \ac{SNN}-based equalizers and demappers outperform the benchmark receivers in terms of \ac{BER}.
Based on these observations, we conclude that \acp{SNN} are powerful models to mitigate the nonlinear distortion caused by the photodiode of \ac{IMDD} systems.
For links experiencing relatively low \ac{CD}, both \nfsnn~and \dfsnn~achieve similar performance. When \ac{ISI} gets more severe, \dfsnn-approaches outperform \nfsnn-approaches, highlighting the importance of decision feedback in such scenarios. \textcolor{Rev2red}{The degraded performance of the \ac{FIR}-based equalizers indicates that the \ac{hcd} link is substantially harder to equalize (Fig.~\ref{fig:lcd_ber} and Fig.~\ref{fig:hcd_ber}). This is physically plausible, as its more than three-fold higher dispersion coefficient causes severe pulse broadening via \ac{ISI} prior to the non-linear photo-detection at the receiver. Nevertheless, by scaling their model complexity ($\ntap$ and $\Nhid$, Tab.~\ref{tab:eq_param}), \ac{SNN}-based equalizers effectively compensate for these non-linearities and maintain low \acp{BER}.}

Tab.~\ref{tab:compare} summarizes the qualitative impact of the different architectural choices on \ac{BER}, $\zavg$, and $\Nt$. \textcolor{Rev1green}{Decision feedback enables better ISI mitigation at high dispersion, while its compact one-hot representation reduces $\Nt$ compared to non-feedback architectures without systematically impacting $\zavg$.
Recurrent connections introduce hidden-layer degrees of freedom that improve equalizer performance (lowering \ac{BER}), but increase both model complexity $\Nt$. 
Furthermore, due to self-excitation of hidden-layer neurons, internal activity $\zavg$ increases.
Finally, activity regularization penalizes excess spiking during training, successfully suppressing $\zavg$ while maintaining low \ac{BER} as long as the loss function remains dominated by the cross-entropy term. Since regularization acts purely on the optimization objective, the architecture and thus $\Nt$ remain unchanged.}
We want to emphasize that regularization effectively counteracts the rise in $\zavg$ when recurrent connections are employed.

Finally, for the investigated values $K = 6$ and $\Nenc = 8$, both encoding schemes \ac{QE} and \ac{RFE} yield comparable performance when applied to the \ac{SNN}-based equalizers and demappers for the investigated links. 
However, when decreasing $K$ below 6, the performance of \ac{RFE} degrades significantly, whereas the performance of \ac{QE} remains approximately stable.
Consequently, for low $K$ and thus low latency, \ac{QE} is the preferred encoding scheme.

\newcommand{\up}{\textcolor{KITred}{\large$\bm{\uparrow}$}}
\newcommand{\down}{\textcolor{KITpalegreen}{\large$\bm{\downarrow}$}}
\newcommand{\neutral}{\textcolor{black}{\large$\text{--}$}}

\begin{table}
    \caption{Impact of architectural choices on performance metrics.}
    \label{tab:compare}
    \centering 
    \begin{tabular*}{.8\columnwidth}{@{\extracolsep{\fill}}lccc}
        \toprule
         & \ac{BER} & $\zavg$ & $\Nt$ \\
         \midrule
         \makecell[l]{Decision feedback} & \down & \neutral & \down \\[1em]
        \makecell[l]{Recurrent connections} & \down & \up & \up \\[1em]
        \makecell[l]{Regularization} & \neutral & \down & \neutral \\ 
        \bottomrule
    \end{tabular*}
\end{table}

We can summarize that the various approaches to the \ac{SNN}-based equalizer and demapper offer a broad range of trade-offs between \ac{BER}, $\zavg$, $K$, and $\Nt$. 
Depending on the specific application requirements, such as low \ac{BER}, a low average number $\zavg$ of spikes per inference, a small number $\Nt$ of trainable parameters, or low latency $K$, the design of the \ac{SNN}-based equalizer and demapper must be adapted accordingly.

\textcolor{Rev1green}{To substantiate the claims of energy efficiency for \ac{SNN}-based equalizers, future work will focus on their end-to-end implementation on custom neuromorphic hardware, such as the SENNA chip~\cite{pscheidl2025senna}, whose ultra-low response latency (down to $20\,\text{ns}$) meets the stringent throughput requirements of high-speed optical communications. A critical consideration in this context is the high-dimensional neural encoding. While currently executed digitally for simulation purposes, a digital encoding pipeline requires high-rate \acp{ADC} and \acp{DAC}, which would introduce significant power overheads and counteract the intrinsic energy benefits of the event-driven equalizer core. To preserve overall system energy efficiency in real-time hardware, this encoding must bypass digital conversion by being realized directly in the analog domain.} \textcolor{Rev2red}{To this end, we believe that \ac{QE} could be efficiently implemented in future hardware via a cascaded array of fast continuous-time comparators. In such a scheme, the analog signal is successively evaluated against halved threshold levels, firing an asynchronous spike and subtracting the threshold value whenever it is exceeded.} \textcolor{Rev1green}{Ultimately, quantifying the exact energy-per-bit savings of the complete receiver pipeline, including the front-end encoding, remains an ongoing research effort that requires physical \acp{IC} integrating both analog encoding front-ends and neuromorphic spiking cores.}

\section{Conclusion}
In this paper, we have presented a systematic comparison of two recently proposed \ac{SNN}-based equalizers and demappers for \ac{IMDD} systems affected by \ac{CD} and nonlinear distortion.
We evaluated the impact of different design choices, including the incorporation of decision feedback, the use of recurrent connections, different neural encodings, and regularization techniques.
Furthermore, we investigated the impact of the encoding parameters on system performance.

Our results demonstrate that the design choices and encoding parameters have varying impacts on the performance metrics, which are the \ac{BER}, average number $\zavg$ of spikes per inference, and total number $\Nt$ of model parameters.
While the incorporation of recurrent connections can significantly improve the \ac{BER}, it also increases $\Nt$ and $\zavg$.
By applying regularization during training, $\zavg$ can be reduced while approximately maintaining the \ac{BER}.
For \ac{IMDD} links with more severe \ac{CD}-induced \ac{ISI}, the incorporation of decision feedback can significantly improve the \ac{BER}.
Our results show that \ac{SNN}-based equalizers and demappers outperforms traditional approaches, offering a highly energy-efficient solution for future \ac{IMDD} systems, with the optimal design depending on application-specific goals, such as low \ac{BER}, minimal spikes per inference, low latency, or reduced model complexity.

%


%




\bibliographystyle{IEEEtran}   
\bibliography{jlt.bib}

\end{document}